# Silicon-based Infrared Metamaterials with Ultra-Sharp Fano Resonances

*Chihhui Wu[1†], Nihal Arju[1], Glen Kelp[1], Jonathan A. Fan[2], Jason Dominguez[4], Edward Gonzales[4,5], Emanuel Tutuc[3], Igal Brener[4,5], and Gennady Shvets[1]**

[1]Department of Physics, The University of Texas at Austin, Austin, Texas 78712; [2]Beckman Institute, University of Illinois at Urbana-Champaign, 405 N Mathews Ave, Urbana, IL 61801; [3]Department of Electrical and Computer Engineering, The University of Texas at Austin, Austin, Texas 78712; [4]Sandia National Laboratories, New Mexico, PO Box 5800, Albuquerque, New Mexico 87185; [5] Center for Integrated Nanotechnologies, PO Box 5800, Albuquerque, New Mexico 87185

ABSTRACT

**Metamaterials and meta-surfaces [ 1, 2, 3] represent a remarkably versatile platform for light manipulation [ 4, 5, 6, 7, 12], biological and chemical sensing [ 8, 9, 10], nonlinear optics [ 11], and even spaser lasing [ 12]. Many of these applications rely on the resonant nature of metamaterials, which is the basis for extreme spectrally selective concentration of optical energy in the near field. The simplicity of free-space light coupling into sharply-resonant meta-surfaces with high resonance quality factors $Q \gg 1$ is a significant practical advantage over the extremely angle-sensitive diffractive structures [ 13] or inherently**

**inhomogeneous high-$Q$ photonic structures such as toroid [14] or photonic crystal [15] microcavities. Such spectral selectivity is presently impossible for the overwhelming majority of metamaterials that are made of metals and suffer from high plasmonic losses. Here, we propose and demonstrate Fano-resonant all-semiconductor optical meta-surfaces supporting optical resonances with quality factors $Q > 100$ that are almost an order of magnitude sharper than those supported by their plasmonic counterparts. These silicon-based structures are shown to be planar chiral, opening exciting possibilities for efficient ultra-thin circular polarizers and narrow-band thermal emitters of circularly polarized radiation.**

The engineering of ultra-sharp metamaterial resonances requires dramatic decrease of the sum total of their respective radiative and non-radiative (Ohmic) losses. One promising approach to decreasing radiative losses while maintaining finite coupling to free-space radiation is to utilize the phenomenon of Fano interference [17] originally introduced in atomic physics to describe asymmetrically shaped ionization spectral lines of atoms. More recently, the concept of Fano resonances was introduced to the field of photonics and metamaterials [18,19,20,21, 22] in which a photonic structure possesses two resonances generally classified as "bright" (i.e. spectrally broad and strongly coupled to incident light) and "dark" (spectrally sharp, with negligible radiative loss). The near-field coupling between the bright and dark resonances leads to coupling of the incident light to the dark resonance which remains high-$Q$. Unfortunately, even for the most judicious engineering of the radiative loss, the total $Q$ is limited [12, 23] by the non-radiative loss of the underlying material. For metamaterials made of metal, Ohmic losses establish an upper limit of $Q \sim 10$, precluding a number of important applications. One such application is the determination of proteins' secondary structure using infrared absorption, which would require mid-infrared ($1500 cm^{-1} < \omega < 1700 cm^{-1}$) metamaterial resonances with $Q \sim 100$ to distinguish between their alpha-helical and beta-sheet conformations [24].

One approach to reducing non-radiative losses is to substitute metallic metamaterials by dielectric ones. Although the electromagnetic properties of dielectric resonators have been studied for decades [25, 26, 27], all-dielectric infrared metamaterials have only recently been demonstrated [28, 29, 30]. Despite this body of work, experimentally demonstrating ultra-sharp metamaterial resonances ($Q \sim 100$) has proven to be challenging, thus greatly impeding further progress in applying metamaterials to practical problems such as biochemical sensing. Here we present an experimental realization of silicon-based infrared meta-surfaces supporting Fano resonances with record-high quality factors $Q > 100$.

The conceptual schematic of the meta-surface used here and an SEM image of a typical sample is shown in Fig.1. The unit cell's geometry is based on the earlier developed plasmonic meta-surface that was utilized for protein monolayer sensing and infrared spectroscopy of single-layer graphene [9, 31]. Each unit cell is comprised of one straight and one bent Si antenna, where the bend is responsible for breaking the two mirror inversion symmetries of the unit cell and coupling the bright (electric dipole) and dark (electric quadripole/magnetic dipole) resonances as schematically shown in Fig.1(c). Hybridization of the two resonances is responsible for ultra-sharp Fano features in transmission and reflection spectra.

In fact, the two antennas comprising the unit cell support several strongly localized dark resonances, as verified using finite-elements methods COMSOL software (Fig.2). All resonances are computed for a meta-surface with the period $P = 2.4\ \mu m$ and antennas' cross-section of $0.5\ \mu m \times 1.2\ \mu m$ in the $x-z$ plane, with remaining dimensions given in the caption. The modes are designated as dark because of their near-vanishing electric dipole moments in the $x-y$ plane and, consequently, weak coupling to the normally incident light. Based on their spatial symmetry, the resonant modes are designated as $TM_{ijk}$, with $i,j,k = 0$ or $1$ corresponding to the $E_y(x,y,z)$ being even or odd under the $x,y,z$ inversion. All dark modes are coupled in the near-field to the bright $TM_{000}$ mode, marked as "dipole" in Fig.1(c). However, because the coupling of the higher-order dark modes to the $TM_{000}$ mode is even weaker than that of the lowest order $TM_{100}$ mode, we expect, and experimentally confirm below, that these modes manifest in even sharper Fano resonances.

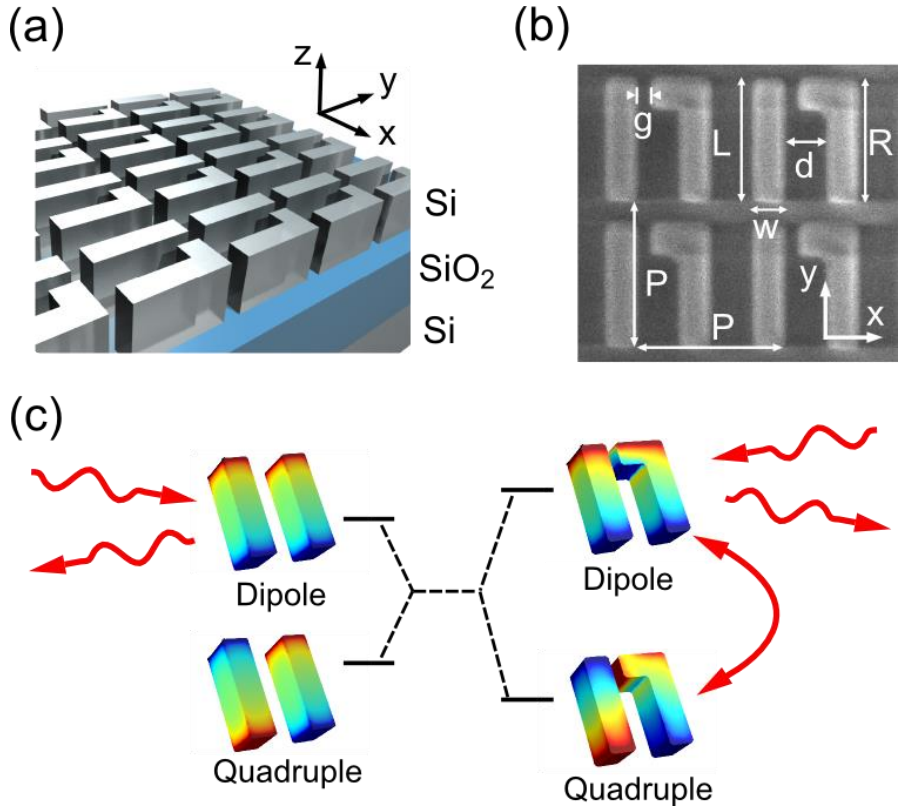

**Figure 1|** (a) A schematic of the silicon-based planar chiral meta-surface supporting ultra-sharp Fano resonances. (b) The SEM image of the fabricated sample and geometry definitions. Physical dimensions: $P = 2.4\ \mu m$, $w = 500\ nm$, $d = 700\ nm$, $g = 200\ nm$, $R = 2\ \mu m$, and $1.6\ \mu m < L < 2\ \mu m$ for the three different samples. (c) A schematic illustrating the Fano interference between electric dipolar (top left) and quadrupolar (bottom left) modes due to the symmetry-breaking small horizontal stub. Calculated charge distributions indicate that the modes approximately retain their spatial symmetry after hybridization.

The experimental and numerical results are presented in Fig.3, where the cross-polarized transmission spectra $T_{ij}(\lambda)$ were acquired using polarized infrared spectroscopy (see Methods for more details) are plotted as a function of the wavelength $\lambda$. The polarizations of the incident/transmitted light ($i, j = x$ or $y$) are set by the polarizer/analyzer, respectively, as shown in Fig.4(a). Spectral tunability of three representative meta-surfaces was accomplished by varying the length $1.6 \, \mu m < L < 2 \, \mu m$ of the left antenna. The $T_{yy}(\lambda)$ spectra provide clear evidence of the Fano interference consistent with Fig.1(c): a broad dip at the frequency of the bright $TM_{000}$ mode at $\lambda_{000} \approx 4.35 \mu m$ is superimposed on a set of narrow features corresponding to the dark modes shown in Fig.2. Similar Fano features are observed in the x-polarized transmission $T_{xx}(\lambda)$, where the broadband background reflectivity originates from the Fabry-Perot substrate resonances.

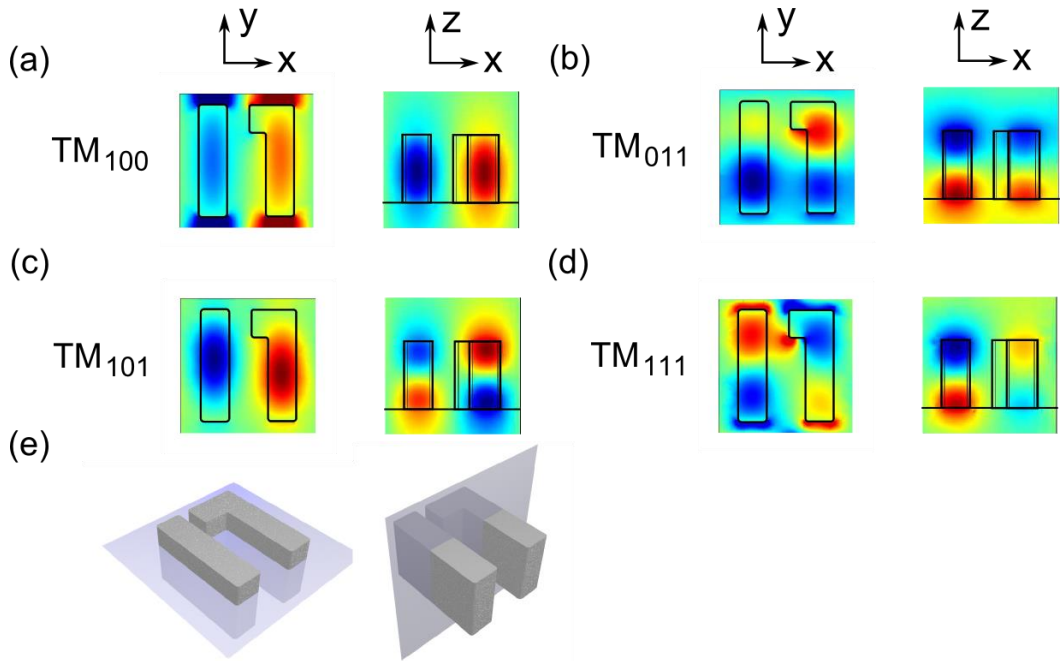

**Figure 2 | Dark resonances supported by the silicon-based meta-surface. (a-d)**: Color maps of $E_y$ in the $x - y$ plane (left) and $x - z$ plane (right). **(e)** Illustration of the cutting planes. The $x - y$ plane is at $z = 800 \, nm$ above the base of the silicon meta-surface. The $x - z$ plane passes through the middle of the unit cell. The corresponding

resonant wavelength are $\lambda_{100} = 4.72\ \mu m$, and $\lambda_{011} = 4.21\ \mu m$, $\lambda_{101} = 4.12\ \mu m$, and $\lambda_{111} = 4.07\ \mu m$, respectively. Physical dimensions of the meta-surface: same as in Fig.1, with $L = 2\ \mu m$.

The most remarkable spectral features are observed in the cross-polarized transmission $T_{xy}(\lambda)$. The baseline $T_{xy}(\lambda)$, small for all non-resonant wavelengths ($\lambda > 5\mu m$), is dramatically peaked at Fano resonances, as shown in Figs.3(c,f), due to the coupling of the dark modes to both $x$ and $y$ polarizations of the incident light. The quality factors $Q = \lambda/\Delta\lambda$ (where $\Delta\lambda$ is full-width half-maximum of each peak) of the Fano resonances, estimated by fitting the experimental cross-polarized spectra with Lorenzian curves, are listed in Table I for the three meta-surfaces.

To our knowledge, these are the narrowest optical resonances observed in mid-IR meta-surfaces. Unlike extremely angle-sensitive diffractive structures [13], Fano-resonant meta-surfaces are ideally matched to far-field radiation with moderate angular divergence $\triangle\theta$ focused by low numerical aperture (NA) optics ($\triangle\theta \approx 7°$ and $NA \approx 0.13$ here). Such experimentally observed angular tolerance translates into minimum acceptable meta-surface size $W_m \sim \lambda/2\triangle\theta$ which can be considerably smaller than $W_d \sim Q\lambda/2$ required for high-$Q$ diffracting structures. Although achieving high-$Q$ resonances depends on collective interactions [33] between neighboring cells of the large area ($300\mu m \times 300\mu m$) meta-surfaces used in our experiments, our simulations confirm that samples as small as $25\mu m \times 25\mu m$ can be utilized without significant deterioration of the spectral sharpness. The unique capability to combine these small area ultra-high $Q$ meta-surfaces with thermal infrared sources is absolutely crucial for the applications described below.

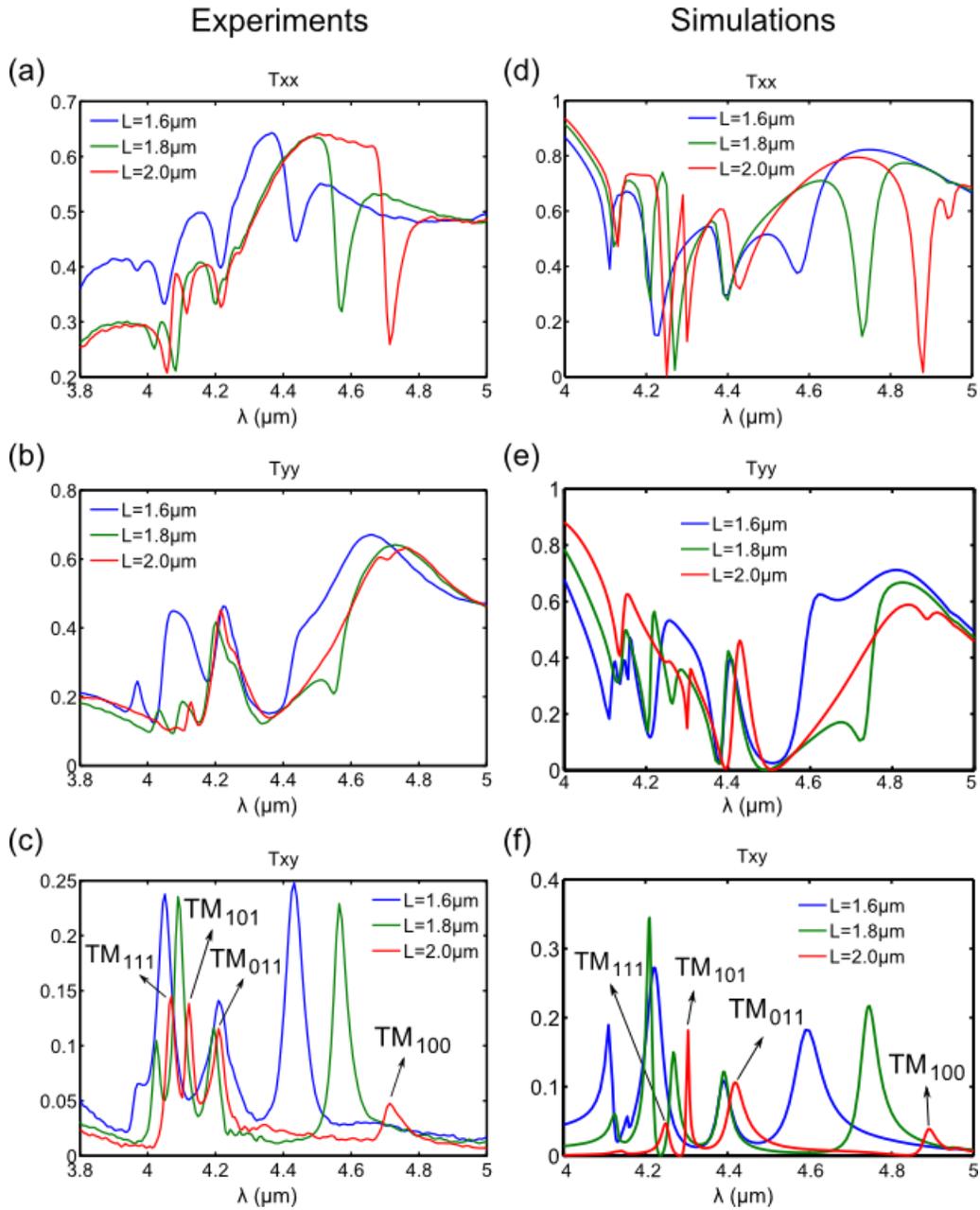

**Figure 3.** Measured (a~c) and calculated (d~f) transmission spectra of the silicon metamaterials with $L = 1.6\ \mu m$ (blue), $1.8\ \mu m$ (green), and $2.0\ \mu m$ (red). The spectra of $T_{xx}$ are shown in (a) and (d), $T_{yy}$ are in (b) and (e), and $T_{xy}$ are in (c) and (f). The four dark resonances are labeled on $T_{xy}$ for the $L = 2\ \mu m$ sample.

| Antenna length → <br><br> Dark mode ↓ | $L = 2\,\mu m$ | $L = 1.8\,\mu m$ | $L = 1.6\,\mu m$ |
|---|---|---|---|
| $TM_{100}$ | $Q = 75.7$, <br><br> $\lambda = 4.71\,\mu m$ | $Q = 94.3$, <br><br> $\lambda = 4.57\,\mu m$ | $Q = 73.9$, <br><br> $\lambda = 4.43\,\mu m$ |
| $TM_{011}$ | $Q = 113.3$, <br><br> $\lambda = 4.21\,\mu m$ | $Q = 100$, <br><br> $\lambda = 4.2\,\mu m$ | $Q = 52.9$, <br><br> $\lambda = 4.21\,\mu m$ |
| $TM_{101}$ | $Q = 113.7$, <br><br> $\lambda = 4.12\,\mu m$ | $Q = 111.7$, <br><br> $\lambda = 4.09\,\mu m$ | $Q = 76.1$, <br><br> $\lambda = 4.05\,\mu m$ |
| $TM_{111}$ | $Q = 116.9$, <br><br> $\lambda = 4.07\,\mu m$ | $Q = 127.5$, <br><br> $\lambda = 4.03\,\mu m$ | $Q = 98.4$, <br><br> $\lambda = 3.97\,\mu m$ |

**Table I | Experimentally obtained quality factors $Q = \lambda/\Delta\lambda$ and spectral positions of the Fano resonances for the three fabricated silicon-based meta-surfaces estimated from fitting the polarization conversion spectrum $T_{xy}(\lambda)$ to Lorenzian shape.**

The first application of the planar (2D) chiral meta-surfaces [ 34 ] described in this Letter, suggested by the high cross-polarized transmission $T_{xy}$, is efficient linear-to-circular polarization (LP-to-CP) conversion. The conversion efficiency and the degree of circular polarization was experimentally investigated using the standard rotating analyzer Stokes polarimetry setup illustrated in Fig. 4(a) to characterize the transmitted polarization state of the $y-$polarized incident light, and to extract its Stokes parameters [ 35] $S_0 = |E_x|^2 + |E_y|^2$, $S_1 = |E_x|^2 - |E_y|^2$, $S_2 = 2\,Re[E_x E_y^*]$, and $|S_3| = -2\,Im[E_x E_y^*]$. A nonzero $S_3$ corresponds to elliptically polarized light, and $S_3 = \pm S_0$ corresponds to right/left CP light. Alternatively, the principal dimensions of the transmitted light's polarization ellipse, its tilt angle $\beta$ and the ratio $a/b$ between its long and short axes defined in Fig. 4(b), can be expressed in terms of the Stokes parameters (see SOM). The measured Stokes parameters and polarization ellipse dimensions for the meta-surface with $L = 1.8 \; \mu m$ are plotted in Fig.4(c,d), and are in good agreement with numerical simulations (see Fig.1S for comparisons and the data for the remaining meta-surfaces).

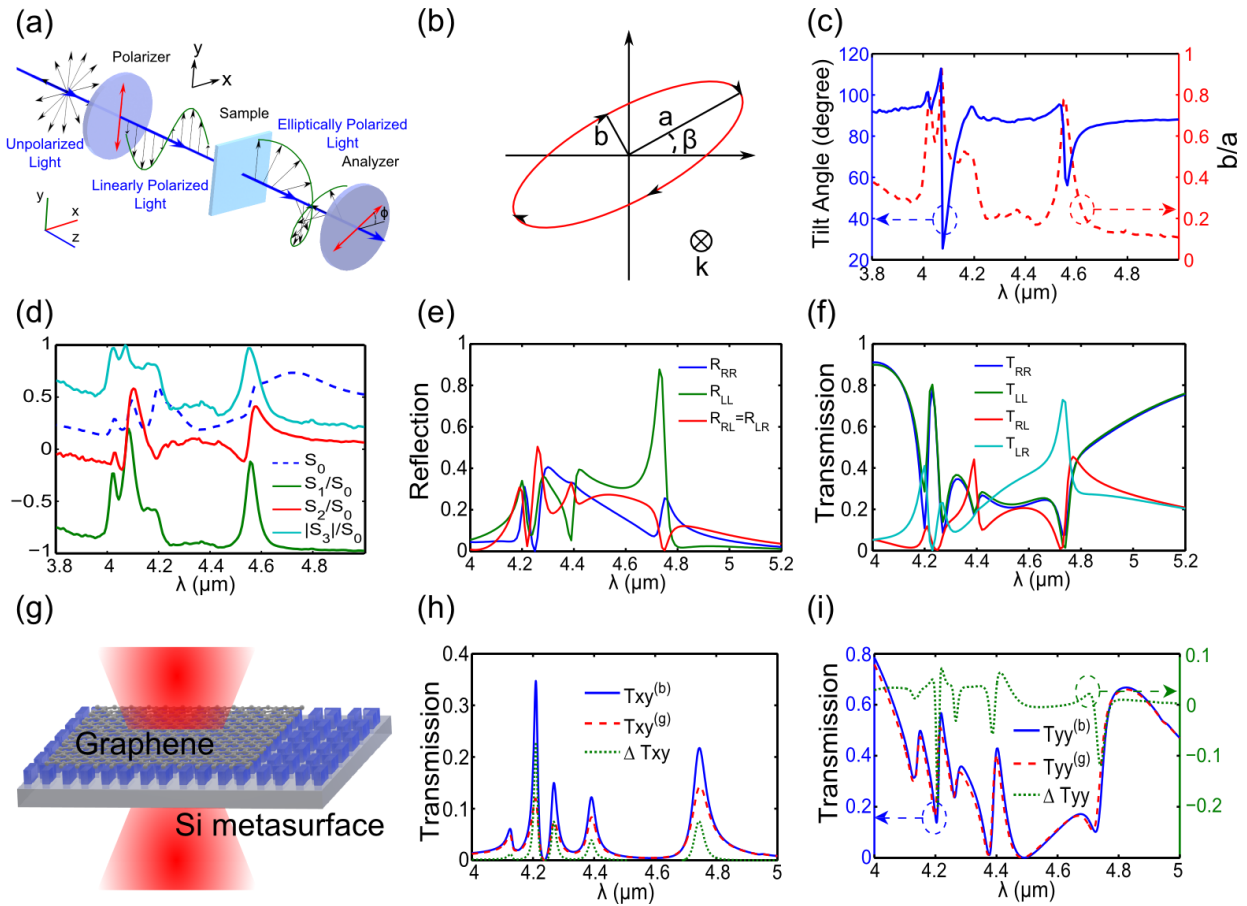

**Figure 4| Applications of Fano-resonant 2D chiral meta-surfaces. (a)** A schematic for the rotating analyzer Stokes polarimetry. A polarizer is used to fully polarize the incident beam in the $y-$direction. **(b)** Parameters definition of the polarization ellipse. **(c)** The measured tilt angle $\beta$ and the inverse of the ellipticity $b/a$ of the polarization ellipse. **(d)** The measured Stokes parameters for the $L = 1.8 \; \mu m$ sample. $S_1$, $S_2$, and $|S_3|$ are normalized with respect to $S_0$. **(e,f)** The calculated reflection/transmission of the CP light for the $L = 1.8 \mu m$ sample. **(g)** A schematic of detecting a single layer of graphene using the silicon meta-surface. **(h)** Cross-polarized transmission $T_{xy}$ before ($T_{xy}^{(b)}$) and after ($T_{xy}^{(g)}$) meta-surface functionalization with single layer graphene with the Fermi level of $E_F$=0 eV. $\Delta T_{xy} = T_{xy}^{(b)} - T_{xy}^{(g)}$ is the transmission reduction due to graphene. (i) Same as (h), but for y-polarized transmission $T_{yy}$ and graphene's Fermi level of $E_F$=0.3 eV.

Note that, away from the Fano resonances, the polarization of the transmitted light is essentially unchanged from its original linear $y$-polarization, as expressed by $S_1/S_0 \approx -1$ in Fig.4(c), and $\beta \approx 90°, b/a \approx 0.1$ in Fig.4(d) for $\lambda > 4.7\mu m$. However, at the Fano resonances the polarization becomes essentially circular, as evidenced by $|S_3|/S_0 \approx 1$ and $b/a \approx 0.8$ at $\lambda_{100} \approx 4.55\mu m$, with conversion efficiency $\approx 50\%$. Even a higher degree of circular polarization ($b/a > 0.9$) is observed for the $TM_{101}$ mode at $\lambda_{101} \approx 4.1\mu m$, thus demonstrating that these meta-surfaces can be used for efficient narrow-band LP-to-CP conversion.

The 2D chiral nature of these meta-surfaces lends itself to another unique application as a source of spectrally-selective CP thermal IR radiation [36] which is uniquely distinct from the non-CP thermal radiation emitted by the natural environment. Even though it is generally assumed that broadband CP emitters are desirable, high spectral selectivity is required for applications such as infrared identifiers (IRID) which rely on unique spectral and polarization signatures of IR tags. Below we briefly outline the conceptual differences between 2D chiral meta-surfaces and other metamaterials used for LP-to-CP conversion, such as the recently demonstrated single-layer plasmonic quarter-wave plates [37] or 3D-chiral metamaterials [7].

Unlike a quarter-wave plate which converts linearly polarized light into either RCP or LCP depending on its initial polarization, the planar chiral meta-surface shown in Fig.1(a) transmits only one CP state. Depending on the position of the antenna bend, the resulting CP state can be engineered to be either RCP or LCP. One of the consequences of the 2D chirality is the strong asymmetric polarization conversion in transmission and reflection near Fano resonances. As shown in Fig.4(f), the RCP-to-LCP and LCP-to-RCP transmission coefficients vary significantly at Fano resonances: $T_{LR} \neq T_{RL}$ despite that $T_{LL} \approx T_{RR}$ as expected for non-3D chiral metamaterials. The implications of strong spectrally-selective reflection asymmetry ($R_{LL} \neq R_{RR}$ as shown in Fig. 4(e)) for applications involving thermal emission of exotic polarization states are even more significant because emissivity is related to reflectivity through Kirhhoff's Law. For example, the circularly polarized emissivity coefficients $\epsilon_R(\lambda)$ and $\epsilon_L(\lambda)$ for a bulk-absorbing emitter are expressed as $\epsilon_R = 1 - R_{RR} - R_{LR}$ and $\epsilon_L = 1 - R_{LL} - R_{RL}$. A simple calculation using the meta-surface reflectivities plotted in Fig. 4(e) shows a high degree of circular polarization $DCP(\lambda) \equiv \epsilon_R/\epsilon_L$ of the thermal emission at the Fano resonance wavelength $\lambda_F \approx 4.7 \mu m$: $DCP(\lambda)$ has a spectral FWHM of $\delta\lambda_{FWHM} \approx 30 nm$ and the peak value of $DCP(\lambda_F) > 20$, which is almost two orders of magnitude higher than its baseline value outside of this narrow resonance region. An additional example of a meta-surface designed to perform as a spectrally-selective thermal emitter of CP IR radiation is shown in Fig.5S.

High quality factor and large optical concentration make the meta-surfaces considered in this Letter an excellent platform for enhancing light-matter interaction in single-layer graphene (SLG) without the need to focus IR light to wavelength-scale spots [ 15]. To demonstrate the effect of SLGs with different free-carrier densities (parameterized by the corresponding values of the Fermi energy $E_F$) placed on top of a meta-surface as shown in Fig. 4(g) on its optical properties, we numerically investigated the induced transmission changes in $T_{xy}$ and $T_{yy}$. The two limiting cases shown in Fig.4(h,i) correspond to $E_F = 0$ (inter-band transitions in SLG make its conductivity real [ 38] and cause light absorption) and $E_F = 0.3eV$ (intra-band transitions cause graphene's inductive response and the blue-shifting of Fano resonances). Significant changes in $T_{xy}$ and $T_{yy}$ near Fano resonances found in both limiting cases demonstrate that silicon meta-surface can be used as a novel platform for spectroscopic characterization of atomic and molecular (e.g., protein) mono-layers.

The ultra-sharp Fano-resonant dielectric meta-surfaces described in this Letter represent a novel and promising platform for a variety of applications that depend on the extremely high optical energy enhancement and precise spectral matching. Those include infrared spectroscopy of biological and chemical substances and nonlinear infrared optics. Chiral properties of such meta-surfaces will be exploited for developing novel ultra-thin infrared detectors sensitive to light's chirality, as well as spectrally-selective CP thermal emitters [36]. Even higher quality factors ($Q > 1,000$) Fano resonant meta-surfaces can be developed by judicious engineering of near-field coupling between resonant modes if inhomogeneous broadening due to fabrication imperfections can be overcome. Combining the extreme field enhancements achieved in such ultra-high $Q$ silicon meta-surfaces with coherent radiation sources such as quantum cascade lasers capable of delivering high-power low-divergence beams [39] would open unprecedented opportunities in nonlinear infrared optics.

## Methods

**Sample preparation.** The samples were fabricated by performing inductively coupled plasma etching on double-layered silicon-on-insulator (SOI) wafers with the thickness of each layer being $h_1 = 1.2\ \mu m$ (Si), $h_2 = 1.6\ \mu m$ (SiO$_2$), $h_3 = 1.2\ \mu m$ (Si), and $h_4 = 1.6\ \mu m$ (SiO$_2$). Only the top active silicon layer is etched into the designed antenna arrays. Below the antenna array are two $1.6\ \mu m$ thick SiO$_2$ layers sandwiching one $1.2\ \mu m$ thick silicon layer, sitting on top of a silicon substrate.

**Optical measurements.** All cross-polarized optical spectra were acquired using a Thermo Scientific Nicolet 6700 FTIR spectrometer. The collimated IR beam emerging from the FTIR spectrometer is passed through the image-forming pinhole, a wire-grid polarizer, focused normally onto the meta-surface using an aspherical ZnSe lens, and then onto an MCT detector placed behind an analyzer. More details can be found in the SOM.

AUTHOR INFORMATION

**Corresponding Author**

Gennady Shvets, gena@physics.utexas.edu.

**Present Addresses**

† Department of Mechanical Engineering , 5130 Etcheverry Hall, University of California Berkeley, CA 94720

**Author Contributions**

C.W. and G.S. proposed the concept. C.W. designed the structures. N.A., J.F., J.D., E.G., E.T., I.B. contributed to fabrication effort. N.A. and G.K. conducted the optical experiments. G.S. supervised the project. The manuscript was written through contributions of all authors. All authors have given approval to the final version of the manuscript.

**Acknowledgements**

This work was supported by the Office of Naval Research (ONR) Grant No. N00014-10-1-0929, National Science Foundation and the NSF award PHY-0851614 This work was performed, in part, at the Center for Integrated Nanotechnologies, a U.S. Department of Energy, Office of Basic Energy Sciences user facility. Sandia National Laboratories is a multi-program laboratory managed and operated by Sandia Corporation, a wholly owned subsidiary of Lockheed Martin Corporation, for the U.S. Department of Energy's National Nuclear Security Administration under contract DE-AC04-94AL85000.